\documentclass[english,submission,copyright,creativecommons]{eptcs}
\usepackage[T1]{fontenc}
\usepackage[latin9]{inputenc}
\usepackage{array}
\usepackage{multirow}
\usepackage{graphicx}

\makeatletter
\providecommand{\tabularnewline}{\\}

\makeatother

\usepackage{babel}
\title{Digital Investigation of Security Attacks on Cardiac Implantable Medical Devices}
\author{Nourhene Ellouze
\institute{Communication Networks and Security Research Lab.\\ University of Carthage, Tunisia.}
\email{nourhene\_ellouze@yahoo.fr}
\and
Slim Rekhis
\institute{Communication Networks and Security Research Lab.\\ University of Carthage, Tunisia.}
\email{slim.rekhis@supcom.tn}
\and
Mohamed Allouche
\institute{Department of Forensic Medicine of Charles Nicolle.\\ Faculty of Medicine of Tunis\\ University of Tunis El Manar, Tunisia.}
\email{mohammad.allouche@yahoo.fr}
\and
Noureddine Boudriga
\institute{Communication Networks and Security Research Lab. \\University of Carthage, Tunisia.}
\email{nab@supcom.tn}
}
\def\titlerunning{Digital Investigation of Security Attacks on Cardiac Implantable Medical Devices}
\def\authorrunning{N. Ellouze, S. Rekhis, M. Allouche, \& N. Boudriga}

\begin{document}

\title{Digital Investigation of Security Attacks on Cardiac Implantable
Medical Devices}
\maketitle
\begin{abstract}
A Cardiac Implantable Medical device (IMD) is a device, which is surgically
implanted into a patient's body, and wirelessly configured using an
external programmer by prescribing physicians and doctors. A set of
lethal attacks targeting these devices can be conducted due to the
use of vulnerable wireless communication and security protocols, and
the lack of security protection mechanisms deployed on IMDs.

In this paper, we propose a system for postmortem analysis of lethal
attack scenarios targeting cardiac IMDs. Such a system reconciles
in the same framework conclusions derived by technical investigators
and deductions generated by pathologists. An inference system integrating
a library of medical rules is used to automatically infer potential
medical scenarios that could have led to the death of a patient. A
Model Checking based formal technique allowing the reconstruction
of potential technical attack scenarios on the IMD, starting from
the collected evidence, is also proposed. A correlation between the
results obtained by the two techniques allows to prove whether a potential
attack scenario is the source of the patient's death.
\end{abstract}

\section{Introduction}

Implantable Medical Devices (IMDs) group together medical devices
that are surgically implanted into patient\textquoteright{}s body
to perform medical therapeutic functions. They are configured using
a programmer which communicates wirelessly with IMDs through the use
of the 402-405 MHz Medical Implant Communication Service (MICS) band
\cite{MICS}. However, the communication protocols used by IMDs to
wirelessly exchange data are insecure. They exhibit a set of security
vulnerabilities, including but not limited to, the use of weak authentication
techniques that are vulnerable to brute-force or replay attacks, the
exchange of clear text or weakly encrypted sensitive medical data
between the IMD and the programmer, and the absence of security mechanisms
to enforce monitoring and detection of unusual behavior or connections.

All of these vulnerabilities make IMDs unprotected and likely to threat
the safety of patients carrying them. Several possible lethal attacks
have been discussed in recent research \cite{zero-power,Design-Challenge-IMD,Sec-IMD,H2H},
including, but are not limited to: a) unauthorized modification of
therapy settings; b) repetitive execution of commands that exhibit
high energy consumption and lead to battery depletion; and c) ordering
the IMD to deliver successive electric shocks leading to fatal arrhythmia.

The existing cardiac IMDs provide a set of digital traces (e.g., electromyogram
history, history of undertaken responses to arrhythmia), which are
useful for conducting a postmortem investigation. They can be interpreted
by pathologists to derive some conclusions about the primary cause
of death. To the best of our knowledge, there are no appropriate mathematical
or formal analysis techniques that were developed to conduct a postmortem
investigation on these traces.

In this context, it becomes necessary to develop new techniques and
methodologies for postmortem investigation of lethal security incidents
on IMDs. The aim is to automatically identify and reconstruct potential
attack scenarios, starting from the evidence collected from IMDs,
and prove the causality between a death of a patient carrying an IMD
and an identified potential attack scenario on such an implanted device.
The need for implementing an investigation technique for IMDs was
discussed in \cite{Sec-Privacy-IMD}. A cryptographic audit log was
proposed to be integrated to an IMD, in addition to the existing logs.
That log allows a secure storage of events related to modification
of IMD settings and provides evidential data for investigators. Authors
in \cite{Cyber-Risk} highlighted also the need for providing and
protecting audit logs. Authors in \cite{BOOK} have highlighted the
need of using an encrypted audit file that logs all access to IMD
settings to identify malicious activities on IMDs. However, to the
best of our knowledge, there is no methodology or technique of attacks
analysis and reconstruction, in the literature, which is appropriate
to the investigation of lethal attacks on IMDs.

Some other works exist in the literature, but are either related to
the investigation of digital attacks on general types of networked
systems, or the investigation of physical crime scenes. Authors in
\cite{carrier2006categories} proposed a digital investigation approach,
which implements a computation model based on a Finite State Machine
and computer history. This approach allows formulating and verifying
hypotheses derived from the occurred events or states in the provided
digital data. Another digital investigation approach addressing the
security attacks on networked systems was presented in \cite{rekhis2011logic}.
The work proposed a formal logic-based approach that allows proving
and reconstructing potential attack scenarios based on a library of
attacks and the provided evidence. In \cite{keppens2006knowledge},
a decision support system allowing the automation of crime scenarios
reconstruction, was proposed. This system, which addresses physic
criminal investigation, is based on a knowledge driven methodology
that instantiates scenarios in forward and backward chaining starting
from a library of stored events. Based on this methodology, a set
of plausible scenarios can be generated according to the provided
evidence. Another physical investigation approach, which is based
on a formal theory of reasoning on evidence, was proposed in \cite{bex2010hybrid}.
It allows the identification of criminal facts by implementing a hybrid
approach of reasoning which integrates the one argument-based approach
and the one story-based approach. This formal theory allows the construction
of hypothetical scenarios in a crime case and the selection of potential
scenarios according to the provided arguments. Authors in \cite{Bayesian-Network}
developed a systematic approach for modeling crime scenarios in a
single Bayesian Network. The latter network, which represents a formal
evidential reasoning based on narrative and probabilistic methods,
enables the development of a software tool improving the communication
between experts and judges in legal cases. An approach to peer review
in forensic pathology is reviewed in \cite{Practice-forensic}. In
this approach, the cause of death is identified, not only through
the results extracted during autopsy, but also through a meeting discussion
that should be performed by pathologists to validate the cause of
death.

In this paper, we propose a digital investigation system for Cardiac
IMDs, which allows identifying and reconstructing potential attack
scenarios that caused a patient's death. A three-step methodology
is proposed, together with a set of techniques to use in each step.
First, the methodology helps identifying the cause of a death based
on the medical observations collected by an IMD, including: a) the
patient\textquoteright{}s electromyogram (EMG) recorded by the IMD
over a period of time before the death; and b) the log file related
to critical reactions undertaken by the IMD (e.g., delivered therapy)
to respond to heart-related emergency situations. An inference system
integrating a set of medical rules is proposed. The latter allows
identifying the potential medical scenarios source of death of a victim
carrying an IMD, through a backward execution of these rules. Second,
based on the access and system logs collected from the IMD under investigation,
which identify the sensitive actions related to access and settings,
we reconstruct the potential technical attack scenarios that would
generate the same logs content if they had been really executed. A
library of attacks is used to support the reconstruction, which is
done in forward chaining. Third, having generated two forms of potential
scenarios: medical and technical, we correlate the two scenarios to
check the existence of an attack scenario that arguments a patient
death.

The paper contribution is four-fold. First, we identify and detail
a set of possible lethal attack scenarios targeting cardiac IMDs,
helping investigators to generate a library of attacks useful for
the reconstruction of technical potential scenarios. Second, compared
to the existing postmortem investigation approaches, which are purely
medical and based on medical observations collected during an autopsy,
our proposal integrates, in addition to the medical forensic postmortem
analysis performed by pathologists, a digital investigation technique
that takes into account the need to identify and reconstruct potential
attacks that may have targeted IMDs. To the best of our knowledge,
we are proposing the first postmortem investigation technique of lethal
attacks on cardiac IDs that integrates the medical and technical aspect
together within a same framework. Third, the whole majority of techniques
described in the proposed system can be automated thanks to the formal
description we have set up, supporting the development of a medical
computer assisted digital investigation tool. Fourth, as the IDs available
in the market show a limitation of storage space and energy resources,
they do not allow to log every action executed by the programmer.
The investigation technique we are proposing allows to infer unobservable
events and generate all potential scenarios that are coherent with
the available evidence, using a library of known attacks.

The remaining part of this paper is organized as follows. The next
section highlights the requirements that should be fulfilled by a
system for postmortem analysis of lethal attacks on IMDs. Section
\ref{sec:Attacks-targeting-Implantable} models and describes attacks
threatening IMDs, stressing on the feasibility of generating a library
of IMD attacks and automating the reconstruction of scenarios. In
Section \ref{sec:Methodology}, we detail the proposed investigation
methodology. Section \ref{sec:Techniques for postmortem  investigation}
describes the techniques proposed to conduct a postmortem investigation
of lethal attacks on IMDs. In section \ref{sec:Case-study}, we provide
a case study, to exemplify the proposal. Section \ref{sec:Conclusion}
concludes the paper.

\section{Requirements for an efficient investigation of attacks on cardiac
IMDs \label{sec:Requirements-for-an}}

A system for postmortem analysis of lethal attacks on Implantable
Medical Devices should fulfill the following requirements.

First, the postmortem analysis techniques should allow differentiating
between natural and criminal death. They should avoid considering
that a death caused by an inappropriate response of an IMD, which
was previously attacked (by modifying its configuration for instance),
is a heart attack related death.

Second, authors in \cite{BOOK} have discussed the importance of developing
an efficient investigation of security attacks and faults on IMDs.
To do so, a set of accurate and reliable copies of evidence should
be generated by the IMDs and made available for investigation. In
this context, at least three types of evidence should be available:
a) traces related to the EMG data collected over a sufficient period
of time preceding the patient's death. They could allow identifying
how and when the patient's health deteriorated; b) traces related
to the IMD responses each time that it detected an emergency situation
(e.g., fibrillation). They would allow identifying inappropriate responses;
and c) traces related to sensitive activities (e.g., authentication,
reconfiguration of therapy parameters) undertaken by remote users
on the IMD. They help identifying potential malicious attacks that
caused the inappropriate responses detected in the second type of
evidence.

Third, as events executed on the IMD are not totally detected and
recorded in the evidential traces provided to investigators (e.g.,
eavesdropping of the exchanged traffic, logs visualization), a postmortem
analysis should hypothesize missing events, leading to the generation
of several potential forms of attack scenarios. Such a generation
could become difficult and complex due to the diversity and complexity
of conducted attack scenarios. Unless a theoretical technique, supporting
the automated inference of scenarios, is developed and used, such
a postmortem analysis could become prone to errors, and sometimes
unable to provide the results within an acceptable time-frame.

Fourth, even if a lethal attack is conducted on an IMD, a technical
investigator cannot solely prove that such an attack is the source
of the patient's death, unless medical experts approve this causality.
On the opposite side, a forensic pathologist cannot prove that a patient
death, which is induced by an inappropriate response of the IMD, can
always be attributed to a criminal modification of the IMD settings.
To overcome this, he can call for technical investigators to help
him demonstrating the occurrence of such malicious events starting
from the collected evidence. Therefore, a system for digital investigation
of lethal attacks on IMDs should be able to reconcile in the same
framework the conclusions derived by technical investigators and the
conclusions generated by medical experts.

\section{Attacks on cardiac Implantable Medical Devices}

\label{sec:Attacks-targeting-Implantable}

IMDs are wirelessly configured through an external programmer, using
in the great majority of cases insecure and vulnerable communication
protocols. These vulnerabilities are mainly related to the use of
very weak authentication mechanisms, and unencrypted or weakly encrypted
sensitive medical data, which can be exploited by executing replay
attacks. In this context, several types of attacks can be launched
on these devices. They can be classified according to their complexity
into: simple attacks and complex attacks.

\subsection{Simple attacks\label{sub:Simple-attacks}}

Simple attacks defines the actions that can be performed by an adversary
during an attack scenario. Below, three types of simple attacks are
described.

\paragraph*{Eavesdropping attacks}

Since the IMD and the programmer exchange data wirelessly, an attacker
can intercept these data and use it for malicious intent. According
to the degree of protection, two types of attacks can be executed.
\begin{itemize}
\item Release of sensitive content: when the exchanged messages are unencrypted,
the adversary analyzes them to discover credentials and extract sensitive
data, including, but not limited to, physiological data provided by
the EMG, therapy configuration, and history of treatments.
\item Traffic analysis: when the exchanged messages are encrypted and secured
using the same cryptographic keys and algorithms, some attacks can
be used in persistent mode (e.g., brute-force attacks) to discover
the credentials used for authentication, and decrypt the exchanged
commands and configurations. The adversary could also try to infer
useful information by observing and analyzing patterns, frequency,
size, and forms of the encrypted traffic.
\end{itemize}

\paragraph*{Unauthorized access to the IMD}

Having discovered the credentials used for authentication, an adversary
can successfully be authenticated to the IMD and thus he can fully
access it and gain privileges of authorized entities (e.g., prescribing
physicians). Therefore, several attacks harming patients\textquoteright{}
safety and IMD security can be executed. These attacks are described
hereinafter.
\begin{itemize}
\item Repetitive electric shocks generation: due to the gained privilege,
the adversary can induce the IMD to deliver a number of successive
electric shocks without detection of any type of arrhythmia.
\item Log modification: an attacker can modify or even delete data recorded
in logs. Such types of attacks are generally used to hide evidence
that help investigators identifying a criminal behavior.
\item Clock drifting: to damage the stored history in an IMD, the attacker
modifies the IMD's clock in such a way that the timestamps in the
event log would be misleading, making events correlation and reconstruction
erroneous.
\item Therapy modification: an adversary alters the therapy settings or
even disables it to, so that the IMD delivers inappropriate therapy
in the future.
\end{itemize}

\paragraph*{Attacking the IMD availability}

Even if an adversary cannot gain access to the IMD, he may damage
the availability of the IMD by executing several types of Denial of
Service attacks. Four types of attacks are described hereinafter.
\begin{itemize}
\item Jamming: by jamming the traffic exchanged between the IMD and the
programmer, a physician would not be able to configure or update the
existing configuration of the therapy to be delivered by the IMD.
In the other side, it may become impossible for the physician to collect
the IMD's logged events using the programmer.
\item Replay: if the exchanged messages are encrypted but not different
from a session to another, the attacker can intercept messages from
one session and replay them later to open a new session as a legitimate
user or re-execute commands. Moreover, by generating a large volume
of replayed messages, an attacker could overwhelm the IMD resources
and drain its battery.
\item Repeated access attempts: by sending access attempts several times,
the IMD battery could be drained. In fact, even if the received messages
are rejected, the IMD loses energy to analyze every received request.
\item Exploiting software vulnerabilities: an authorized entity, who gained
access to an IMD, can remotely update the IMD's software (this feature
was discussed in \cite{BOOK}). Therefore, an adversary can analyze
the firmware of the IMD and identify the remote vulnerabilities by
which he can send counterfeit updates or firmware, leading the IMD
to not react suitable during emergency.
\end{itemize}

\subsection{Complex attacks\label{sub:Lethal-attacks}}

\begin{figure*}
\begin{centering}
\includegraphics[scale=0.3]{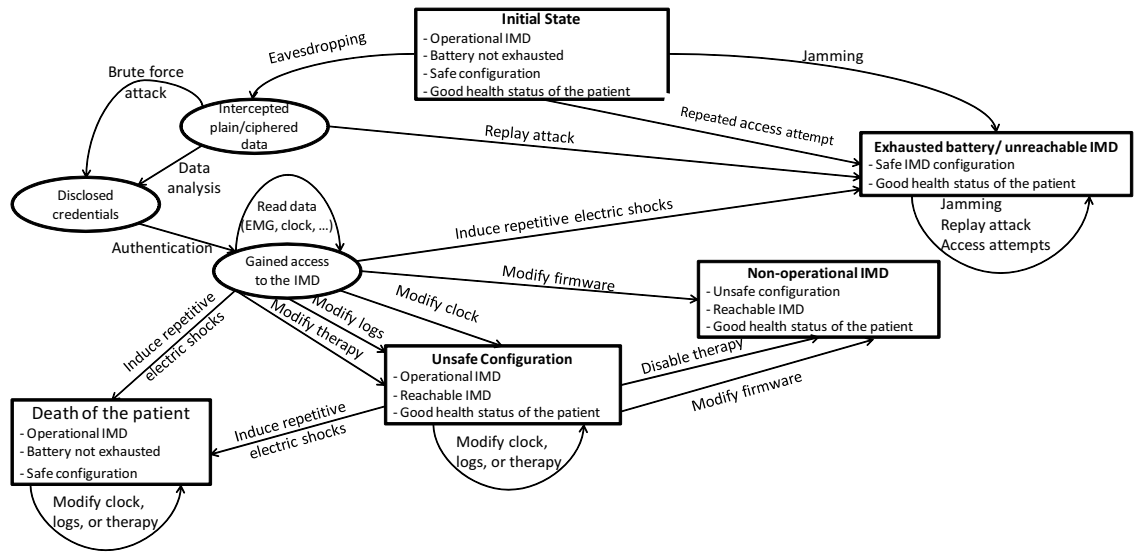}
\par\end{centering}

\caption{Lethal attack scenarios\label{fig:Lethal-attacks-scenarios}}
\end{figure*}

By combining several forms of simple attacks, as described in the
previous subsection, complex attack scenarios can be generated. Some
of these attacks could be lethal. These scenarios are depicted in
the graph provided in Figure \ref{fig:Lethal-attacks-scenarios}.
Every node in this graph specifies the state of the IMD seen throughout
the execution of the attack (e.g., connected users, configuration
data, and available quantity) and the health status of the patient
carrying this IMD. These states can be classified into secure and
insecure starting from a list of invariants that can be generated
by experts. Two states $s_{1}$ and $s_{2}$ in the graph are linked
together, if an action $A$ is executed from $s_{1}$ leading to the
generation of $s_{2}$. To be executed action the $A$ should be enabled
in the state $s_{1}$. The automated generation of such a graph could
be done starting from a library of simple attacks and an initial description
of the initial state of the IMD (i.e., the state at which the IMD
started working, or the state identified by the physician during the
last medical examination to be a normal state).

We describe in the following the typical anatomy of a complex attack
scenario on an IMD. Generally, an adversary starts by intercepting
and analyzing the exchanged traffic between the IMD and the programmer
to discover credentials used for authentication. Upon discovering
them, he tries to gain access to the IMD by proceeding through authentication.
Having gained access, he becomes able to execute different type of
attacks including: a) the execution of inappropriate actions such
as forcing the IMD to deliver electric shocks; b) the modification
of the configuration (e.g., clock drifting, logs alteration, or therapy
settings modification); and c) the deactivation of the life-sustaining
functions (e.g., updating the firmware, disabling the therapy). A
wide set of examples of lethal attack scenarios can be defined. Among
the most important, we present the following.

Thanks to the use of MICS band, a transmission distance of 10 meters
can be achieved between an IMD and a programmer. An eavesdropper located
within this transmission range can intercept the exchanged traffic
between the IMD and the programmer. Based on a known specification
of the IMD's radio chip and using a software radio platform, the eavesdropper
can also conduct a reverse engineering on the protocol and packet
structure used by the IMD to communicate with the programmer. Consequently,
as existing IMDs were not made secure for the sake of reduction of
energy overhead, an attacker would be able to discover the used credentials
and also to forge packets containing erroneous configuration parameters.
An example of this attack was shown in \cite{Pompeinsuline} over
insulin pumps. Moreover, in \cite{zero-power}, a reverse-engineering
of the communication protocol used by an Implantable Cardioverter
Defibrillator (ICD) was conducted using an oscilloscope and a Universal
Software Radio Peripheral (USRP). A set of software radio-based attacks,
including but are not limited to, replaying of identification data
exchanged between the ICD and the programmer during access, modification
of the ICD's clock, and update of the therapy settings, was conducted.

\paragraph*{Scenario 1}

An attacker intercepts the traffic (we suppose that messages are unencrypted)
exchanged between the IMD and the programmer, and analyzes it to extract
credentials and data related to the patient's health status and therapy
settings. After being successfully authenticated to the IMD, he modifies
the therapy settings, based on the collected information, in such
a way that the new configuration would be unsafe to the patient health.
A future occurrence of some types of arrhythmia would be fatal, as
the IMD could respond inappropriately.

A malicious modification of the therapy settings, could consist in
replacing the thresholds associated to the detection of Ventricular
Tachycardia ($VT$), so that a future occurrence of an Atrial Fibrillation
($AF$) will be detected by the IMD as a $VT$ (An example of an inappropriate
detection of $AF$ as $VT$ is shown in \cite{Proarhythmia}). To
respond, the IMD will deliver inappropriate responses (e.g., antitachycardia
pacing), leading to the occurrence of successive proarrhythmia events
(e.g., $VT$ events). For example, in \cite{Proarhythmia}, when the
IMD wrongly detects a $VT$, it delivered a 5.1-J cardioversion shock
followed by a 35.1-J defibrillation shock. Faced to the latter response,
the patient's heart starts to experience a syncopal $VT$, leading
the IMD to deliver a new 35.1-J defibrillation shock. This response
induces a faster $VT$, leading the IMD to deliver a 34.4-J shock,
which cause a posterior deceleration. In this scenario, the IMD has
exhausted the maximum energy shocks without correcting the occurred
$VT$. Unless, a precordial thump, is performed by healthcare professional
to correct this arrhythmia, the patient could die.

\paragraph*{Scenario 2}

In this scenario, we suppose that the exchanged traffic between the
IMD and the programmer is encrypted. Therefore, even by intercepting
this traffic, the adversary would not be able to discover credentials.
If the used cryptographic algorithm and keys are kept unchanged during
the whole lifetime of the IMD, it may be possible for the attacker
to conduct a brute force attack in persistent mode and get the IMD
security credentials. After discovering these credentials and being
successfully authenticated to the IMD, he sends commands to the IMD
to deliver successive electric shocks, leading to the acceleration
of the patient's heartbeat which could cause fatal arrhythmia.

\paragraph*{Scenario 3}

As detailed in the previous scenario, and after the attacker gains
access to the IMD, he disables the life-sustaining functions of the
IMD by disabling the configured therapy. After the IMD becomes inoperative,
the occurrence of arrhythmia could cause the patient's death.

While existing IMDs are not designed to support digital investigation,
some of the traces they provide could be useful for investigation.
In fact, by correlating sensitive data collected from logs (e.g.,
modification of therapy parameters, time and nature of responses undertaken
by the IMD) with the EMG recorded by the IMD, an investigator would
be able to deduce some lethal attacks. For instance, some actions
in the second scenario could be detected after noticing that several
electric shocks were delivered without any occurrence of arrhythmia.

\section{A methodology for postmortem investigation of attacks on cardiac
IMDs \label{sec:Methodology}}

In this section, we provide an overview of the evidence that can be
collected from IMDs during a postmortem investigation, and describe
the proposed methodology. A tamper-evident logging mechanism is supposed
to exist on the IMD being investigated, so that the integrity of evidences
can be checked before analysis. In this context, while an eavesdropper
can a) capture the exchanged traffic and collect credentials, which
are used for authenticating a privileged user (e.g., physician) to
the IMD; and b) authenticate himself from an unauthorized programmer
using the discovered authentication credentials; and c) gain a privileged
access on the IMD; he cannot alter or modify any log file generated
on this IMD. An additional layer of security can be used to guarantee
that access to IMD's logs is only done by an authorized physician,
who can read its content and decide to reduce the size of that file
(if it is close the maximum value) by deleting old and low-severity
events generated previously to the last medical consultations. While
a Write Only Memory (WOM) can not be used as a solution as the physician
should be able to read the log history using the same device components,
the mutual authentication protocol, which was proposed in \cite{Trusted2013},
could be applied in such cases to enforce authentication. A set of
cardiac sensors could be connected during remote access to this IMD
logs to collect the patient's electrocardiogram and extract a biometric
key. During the same time the IMD is asked to extract a biometric
key from the EMG that it is monitoring. If the generated key is the
same, then authentication is successful.

\subsection{Evidence description}

Postmortem investigation of attacks targeting Implantable Medical
Devices is performed based on a set of evidence, which can be classified
into two technical evidence and medical evidence.

\paragraph*{Technical evidence}

They provide a history of the set of sensitive events that occurred
on the IMD configuration or triggered by the connected users. Examples
of provided events include, but are not limited to, successful and
unsuccessful attempts of remote access, modification of the IMD\textquoteright{}s
clock value, and alteration of the therapy or firmware. Such a history
is typically collected from the IMD log as a series of time stamped
events. It may provide an incomplete and limited description of an
attack scenario conducted by an attacker before the patient's death,
or before inducing the IMD to a denial of service.

\paragraph*{Medical evidence}

They are a set of findings derived by pathologists during investigation
based on the examination of the victim and the carried IMD. These
evidence represent: a) the patient\textquoteright{}s electromyogram
(EMG) which was continually recorded by the IMD over the period of
time before the death. Since the EMG represents the recording of electrical
activity of heart muscles, such an evidence allows to identify a wide
set of cardiac anomalies through the observation of the electrical
waves (e.g., P wave, T wave, and R wave) related to the heart activity
of the patient before he died; b) the set of reactions (i.e., the
therapy delivery) undertaken by the IMD over time. An IMD reaction
is an electric shock that is delivered to correct a detected arrhythmia.
The value of this electric shock, which depends on the type of detected
arrhythmia, is expressed in Joule. These responses, which are time-stamped,
can be matched to the anomalies identified on the EMG; c) the set
of clinical observations collected during autopsy, such as the status
of the probe of the implanted device; and d) the set of additional
information collected from the medical record of the dead person (e.g.,
chronic illness, medications).

\subsection{Three-step evidences analysis}

We describe in this subsection a three-step methodology of postmortem
investigation of lethal attacks on cardiac IMDs, operated by a group
of pathologists and technical investigators.

\paragraph*{Medical investigation}

This step aims at conducting a medical investigation based on the
medical evidence (as defined in the previous subsection) collected
from the IMD under investigation. A set of invariants, describing
anomalies that a physician can detect on an EMG and a set of unexpected
responses of the IMD under investigation when these anomalies occur,
are defined. These invariants allow to transform an EMG record, together
with the traces describing the reactions undertaken by an IMD, into
a medical timestamped log containing a description of the different
forms of sensitive heart-related events (e.g., ventricular fibrillation
and ventricular extrasystole) and the reaction undertaken by the IMD,
if any, for each one of them.

A set of inference rules is proposed to describe causal relations
between the different types of arrhythmia and IMD reactions, showing
all premisses of each arrhythmia. These rules constitute an inference
engine that is executed in backward chaining, starting from the heart
death event, allowing pathologists to automate the generation of potential
medical scenarios. Each scenario describes an ordered list of dependent
arrhythmia and IMD responses, showing the evolution of the health
status of the patient until he died. A generated medical scenario
can, for example, demonstrate that an arrhythmia occurred or was amplified
due to an inappropriate response of the IMD.

\paragraph*{Technical investigation}

During this step, attacks scenarios on the investigated IMD are reconstructed
starting from the collected technical evidence. A description of the
initial system state of the IMD is generated by technical investigators,
using an investigation library previously provided. The latter contains
a description of all simple attacks on IMDs and the actions executed
by the IMD in response to remote users\textquoteright{} actions or
detected arrhythmia. A Model Checking based algorithm is developed
in this paper to automate the generation of potential attack scenarios
starting from that library and a description of the functions by which
an IMD generates logs. A potential scenario generated by this Model
Checker will represent a series of actions (retrieved from the investigation
library), which, if combined and executed, would produce evidence
that are similar to those collected from the IMD. To be considered
as malicious, an attack scenario should contain at least one malicious
action Contrarily to the Model Checkers existing in the literature,
the one we are discussing in this paper is dedicated for the forensic
investigation. It generates scenarios in conformance with the collected
evidence taking into consideration two categories of actions: visible
and invisible.

\paragraph*{Correlating reconstructed medical and technical scenarios}

While the second step allows to generate potential attack scenarios,
it cannot be used to state that these scenarios are responsible of
the patient\textquoteright{}s death. The objective of this step is
to correlate the scenarios obtained by medical investigation and the
scenarios obtained by technical investigation. The aim is to prove
that the malicious actions in the technical scenarios of attacks caused
the absence of reactions or the inappropriate reactions as shown in
the medical scenarios.

\section{Techniques for postmortem investigation}

\label{sec:Techniques for postmortem  investigation}

In this section, we describe for each step of the investigation methodology
the techniques to use.

\subsection{Medical inference system\label{sub:Inference system rules}}

In the following, we describe an inference system for medical investigation.
It uses the set of medical inference rules, discussed in the previous
subsection. These rules describe the causal relations between events
denoting types of arrhythmia and IMD reactions. An inference rule
takes the following form $\left(Ev\right)^{n}\rightarrow_{T}\,\left(Ev'\right)^{m}$,
where $n$ and $m$ are positive non-zero integers. That rule means
that the occurrence of $Ev$ for $n$ times successively leads to
the occurrence of $Ev'$ $m$ times successively within a period of
time $T$. $Event$ described in such a rule is a conjunction and/or
disjunction of simple events that can be classified into observable
and unobservable events. Observable events are events that can be
detected by the IMD under investigation and included in the collected
medical evidence (e.g., ventricular fibrillation, electric shocks).
Unobservable events are events that are not detected by the IMD sensors,
and therefore cannot be included in the collected medical evidence
(e.g., Acute Pulmonary Edema).

Due to the complexity of human body, a wide set of heart reactions,
that are useful for the automated generation of medical scenarios,
could be identified and used to develop a large set of inference rules.
For the sake of space, we have only selected a non exhaustive set
of $12$ rules (rules $1$ to $11$ require the occurrence of events
one time only $\left(m=n=1\right)$) described in Table \ref{tab:Example-of-medical-rules}.

\begin{table}
\begin{centering}
\begin{tabular}{|>{\raggedright}p{4cm}|>{\raggedright}m{10cm}|}
\hline
\multirow{1}{4cm}{Medical rules} & Descriptions\tabularnewline
\hline
\hline
$(1)\, VF,\, AR\rightarrow_{T}\, VF$ \\
 & \multirow{2}{10cm}{Rules $(1)$ and $(2)$ state that the occurrence of a Ventricular
Fibrillation ($VF$), to which the IMD does not respond ($AR$) or
responds inappropriately ($IR$), leads to the occurrence of another
episode of $VF$ within the following period of time of length $T$}\tabularnewline
\cline{1-1}
$(2)\, VF,\, IR\rightarrow_{T}\, VF$\\
 & \tabularnewline
\hline
$(3)\, VF,\, AR\rightarrow_{T}\, HD$\\
 & \multirow{2}{10cm}{Rules $(3)$ and $(4)$ are similar to rules $(1)$ and $(2)$, in
the sense that they use the same premises, but generate a consequence
$HD$ instead of $VF$.}\tabularnewline
\cline{1-1}
$(4)\, VF,\, IR\rightarrow_{T}\, HD$\\
 & \tabularnewline
\hline
$(5)\, VES,\, AR\rightarrow_{T}\, VF$\\
 & \multirow{2}{10cm}{Rules $(5)$ and $(6)$ state that the occurrence of a Ventricular
Extrasystole ($VES$), to which the IMD does not respond, or responds
inappropriately, leads to the occurrence of a $VF$, within a period
of time of length $T$.}\tabularnewline
\cline{1-1}
$(6)\, VES,\, IR\rightarrow_{T}\, VF$\\
 & \tabularnewline
\hline
$(7)\, VT,\, AR\rightarrow_{T}\, VF$\\
 & \multirow{2}{10cm}{Rules $(7)$ and $(8)$ are similar to rules $(5)$ and $(6)$, respectively,
in the sense that the same consequences are generated using the premise
Ventricular Tachycardia ($VT$) instead of $VES$.}\tabularnewline
\cline{1-1}
$(8)\, VT,\, IR\rightarrow_{T}\, VF$\\
 & \tabularnewline
\hline
$(9)\, VT,\, AR\rightarrow_{T}\, HD$\\
 & \multirow{2}{10cm}{Rules $(9)$, and $(10)$ show for the same premises occurring in
rules $(7)$ and $(8)$, that $HD$ can be another consequence.}\tabularnewline
\cline{1-1}
$(10)\, VT,\, IR\rightarrow_{T}\, HD$\\
 & \tabularnewline
\hline
$(11)\, ST,\, IR\rightarrow_{T}\, ST$\\
 & Rules $(11)$ states that the occurrence of a Sinus-Tachycardia ($ST)$,
to which the IMD does not respond ($AR$), leads to the occurrence
of another episode of $ST$, within the following period of time of
length $T$.\tabularnewline
\hline
$(12)\,\left(ST,\, IR\right)^{n}\rightarrow_{T}\, VF$\\
 & Rules $(12)$ states that the repetitive occurrence of Sinus-Tachycardia
($ST$), followed by an inappropriate response $(IR$) of the IMD,
leads to the occurrence of a $VF$, within the following period of
time of length $T$.\tabularnewline
\hline
\end{tabular}
\par\end{centering}

\caption{Example of medical rules and their description \label{tab:Example-of-medical-rules}}
\end{table}

\subsection{Inferring medical scenarios}

In order to infer the potential medical scenarios, pathologists execute
the inference rules in backward chaining starting from an observed
heart death, which represents the last observed event in the medical
evidence, and the collected medical evidence. Several potential medical
scenarios can be generated, creating a tree whose root node corresponds
to an $HD$ event. The tree of medical scenarios will be reconstructed
as follows. Assuming that $Ev$ is an event in the reconstructed tree
of medical scenarios, and $Ev'$ is the last observable event in the
scenario connecting $DH$ to $Ev$ in the tree (if $Ev$ is observable
then $Ev'$ is equal to $Ev$), a rule in the form of $E\rightarrow_{T}\, E'$
can be executed in backward chaining, unless the following two conditions
are met:
\begin{itemize}
\item The consequence $E'$ in the rule corresponds to event $Ev$.
\item If the premiss $E$ is observable, it should correspond to the event
immediately preceding $Ev'$ in the medical evidence, and occurring
no earlier than $T$.
\end{itemize}
Once the rule is executed, event $E$ is appended to the tree under
construction and $E'$ is linked to $E$.

This process of reconstructions stops when: a) none of the inference
rules can be executed; b) the events in the reconstructed graph start
to be older than a predefined threshold duration; and c) all recent
arrhythmia events in the medical evidence have been included in the
tree.

\subsection{Technical investigation}

To reconstruct technical attack scenarios, investigators combine and
execute actions from the library of attacks to generate scenarios
that could provide the same collected evidence. We describe in the
following the mechanisms by which a scenario is reconstructed. We
start by formalizing the description of an attack scenario, say $\omega$,
that we model in the form of $\omega=\langle s_{0},\, A_{1},$ $s_{1},\,...,\, s_{n-1},$
$A_{n},\, s_{n}\rangle$, where $s_{0}$ is the initial system state,
$A_{i}$ ($i\in[1.n]$) represents elementary actions executed in
the scenarios, and $s_{i}$ ($i\in[0..n]$) is a description of the
intermediate system state generated after the execution of an action
$A_{i}$. To be executed, an action $A_{i}$ should be enabled at
state $s_{i-1}$, and once executed it generates state $s_{i}$. We
note that an action $A$ in $\omega$ could be legitimate or malicious.

We denote by $obs(\,)$ an observation function over states, actions,
and attack scenarios, which can be used to describe any form of monitoring
and security supervision mechanism deployed on the IMD. The output
of such a function denotes the observable part of action, state, or
a scenario. It describes the evidence that is typically generated
further to noticing a new state, action, or scenario. With respect
to a security solution, an action or a state could be: a) Visible,
in the sense that its value can be captured by the observer; or b)
Invisible, in the sense that none information regarding its occurrence
could be determined.

To compute $obs(\omega)$, first, we set $\overline{obs(\omega)}$
as the sequence $\langle obs(A_{1}),$ $...,obs(A_{n})\rangle,$ after
replacing each state action $A_{i}$ by $obs(A_{i})$ and eliminating
all states (we suppose that the security mechanisms deployed on the
IMD are event-based, and therefore only able to monitor events instead
of supervising states). Second, we set $obs(\omega)$ equal to $\overline{obs(\omega)}$
after eliminating unobservable actions. We can notice that a security
solution is unable to log unobservable events nor detect their occurrence.
From the investigation point of view, two consecutive events in the
generated evidence could be considered as two adjacent events in the
reconstructed scenario, or be separated by a series of unobservable
events that could be retrieved from the library of investigation.

To reconstruct attack scenarios, a Model Checking based algorithm
is executed in forward chaining as follows. Starting with a description
of the IMD's initial system state (supposed to be known in advance,
as we stated previously), the algorithm retrieves actions from the
library and executes them to generate potential scenarios that would
produce the same provided evidence. Given a scenario $S=\langle s_{0},A_{1},s_{1},...,A_{i},s_{i}\rangle$
under construction, such that $obs(S)\sqsubseteq E$ (i.e., the evidence
expected to be generated by the IMD further to the occurrence of the
scenario $S$ is part of $E$, starting from the beginning of file),
the algorithm proceeds as follows. It checks whether there is an action
$A$ in the investigation library such that $A$ is enabled in $s_{i}$.
If yes, it executes $A$ starting from state $s_{i}$ to generate
state $s'$ ($s'=A(s)$). Once $s'$ is generated, the algorithm checks
if the evidence expected to be generate by the IMD, if the scenario
$\omega'=\omega|\langle A,s'\rangle$ is executed, is part of $E$.
If yes, it sets the scenario $\omega$ equal to $\omega|\langle A,s'\rangle$.
As this algorithm uses the Model Checking technique, it tries to execute
any possible action from the library of investigation (as we described
above), leading to the generation of a graph of technical scenarios
satisfying the provided evidence. Since the generation of scenarios
is done with respect to the set of collected evidences, the problem
of state explosion is highly reduced since an observable action cannot
be selected from the library unless it is satisfied by the collected
evidence (which contains also a finite set of observed events and
states).

\subsection{Correlating potential scenarios}

The two previous described techniques, i.e., the inference rules and
the Model Checking, allow only the generation of medical scenarios,
and potential technical scenarios, respectively. Nevertheless, they
do not allow to prove at least one of the generated potential medical
scenario is caused by one of the generated potential technical scenarios,
unless the two scenarios are correlated together. In fact, it may
happen that a medical scenario that lead to the death of the patient
was caused by a displacement of IMD leads (leading the IMD response
to look inappropriate) and not by a malicious attack on the IMD. The
aim here is to show how correlation can be performed.

The opinion of the pathologist is important to validate the generated
medical scenarios, and the performed correlation between medical and
technique scenarios, due to the following reasons. First, the generated
medical scenarios could contain unobservable events by the IMD. Second,
patients are different from each other and some of the generated medical
scenarios could be unrealistic.

Correlation of scenarios is performed as follows. First, a potential
medical scenario is examined to check the existence of suspicious
IMD responses (e.g., $IR$ and/or $AR$), and identify the set of
parameters related to that response in the medical evidence (e.g.,
heart rate, arrhythmia duration, EMG amplitude). Second, potential
technical scenarios are analyzed to identify malicious actions threatening
the security of the IMD, and identify the modifications brought by
them on the configuration, firmware, or battery energy. Third, a correlation
is performed to check whether one of the suspicious IMD responses
identified in the first step is caused by one of the malicious actions
identified in the second step. If yes, it can be proved that a potential
attack scenario on the IMD under investigation was conducted and caused
the identified medical scenario (that lead to the patient death).

One example of correlation is described as follows. When examining
the generated potential medical scenario, a pathologist notices the
existence of a set of IMD's inappropriate responses (i.e., $IR$).
In the other side, the generated potential technical scenario showed
a modification on the therapy settings. Therefore, the pathologist
decides to deeply analyze the provided technical evidence, and examine
the updated therapy parameters and their values. If he notices that
the non-application of these updates would have prevented the occurrence
of the set of observed inappropriate responses, and thus the death
of the patient, then he can prove that the generated potential technical
attack scenario conducted on the IMD lead to the generated medical
scenarios.

\section{Case study}

\label{sec:Case-study}

To exemplify the proposal, we present a case study related to the
investigation of a lethal attack on an IMD, where the results obtained
at each step of the methodology are detailed and discussed. The investigated
IMD is supposed to use an authentication protocol which is vulnerable
to replay attacks. It was impossible for us to find a real case study
where a lethal attack was conducted on a patient's IMD. For that reason,
the case study in this section was inspired from \cite{Exple-Scenario},
which shows how a noise over-sensing in an IMD can create adverse
effect on health outcomes and mortality. We have introduced some modifications
to that scenario by replacing the noise over-sensing by the execution
of an unauthorized therapy modification on the IMD. It is important
to mention that the analysis proposed in \cite{Exple-Scenario} does
not allow to cope with security issues.

\subsection{Description of the conducted scenario}

\paragraph{Description of the criminal attack scenario}

An adversary, who was able to acquire credentials used to authenticate
himself to an IMD, sent an authentication request to the IMD and succeeds
in gaining access. After that, he read the the patient's health status,
by examining the data stored in the IMD log. Then, he modified the
therapy settings, especially by altering the thresholds associated
to the detection of some arrhythmia. The IMD becomes unable to appropriately
detect, identify, and respond to arrhythmia. Later, the attacker disconnected.

\paragraph{Description of the medical incident}

Few hours after, the patient gets a Sinus-Tachycardia ($ST$). Since
the IMD is misconfigured, such an event is detected as a Ventricular
Fibrillation ($VF$). The IMD responds by delivering a significant
electric shock. Due to the continuous inappropriate detection, the
same arrhythmia continues to happen and five other electric shocks
were delivered additionally, among them, the last one led to the occurrence
of a true $VF$.

Faced to that $VF$, the IMD could not respond, since the maximum
number of electric shocks (set by the physician to $6$), that can
be delivered within a predefined period of time, has been reached.
During the consequent deactivation period successive episodes of $VF$
occurred without being responded by the IMD, leading to the patient
death.

\subsection{Postmortem analysis}

During the postmortem investigation process, the set of techniques,
which we described in Section \ref{sec:Techniques for postmortem  investigation},
are executed.

\paragraph*{Medical and technical evidence collection}

Collection of technical evidence from the IMD revealed four timestamped
events that we describe in their order of appearance: $T1$) Session
opening further to a successful authentication (user id and session
identity are provided); $T2$) modification of therapy settings (new
updated values are provided); and $T3$) session closing (session
identity is provided).

\paragraph*{EMG history and IMD responses traces}

Collection of medical evidence revealed the following events related
to important arrhythmia and IMD responses, that we describe in their
order of appearance: $M1$) six successive episodes of $ST$, each
one of them is immediately followed by a delivery of a therapy suitable
for$VF$; $M2$) several episodes of $VF$, whose intensity grew from
an episode to another. None of these instances of $VF$ is followed
by an IMD response (this corresponds to event $AR$); and $M3$) heart
death.

\paragraph*{Generation of potential medical scenarios}

Starting from the collected medical evidence, and based on the set
of inference rules described in Subsection \ref{sub:Inference system rules},
we generate one potential medical scenario as follows. Starting with
the last event $M3$, which represents a $HD$, rule $(3)$ can be
executed in backward chaining, as the premisses of the rule correspond
to $M2$. Later rule $(1)$ is executed in backward chaining a number
of times equal to the instances of $VF$ in event $M2$. Finally rule
$(12)$ is executed as its consequence correspond to $VF$ and its
premisses are provided by $M1$. The output of the used inference
system is a medical scenario in the form of $(M1,\, M2,\, M3)$.

\paragraph*{Generation of technical attack scenarios}

Based on the provided traces collected from the IMD log and the investigation
library describing simple attacks targeting IMDs, two potential attack
scenarios are identified as follows.

In the first scenario, say $S1$, the attacker eavesdrop the credentials
(action $A1$) exchanged through messages that are weakly encrypted
due to energy issues (use of a weak and lightweight encryption algorithm,
use of static keys). By brute-forcing these credentials, he authenticates
himself to the IMD (action $A2$). After successfully authenticating
the attacker, the IMD opens a session (action $A3$). Then, the attacker
reads the stored medical data (action $A4$) (e.g., EMG history, therapy
settings). Later, he alters the therapy settings (action $A5$), and
disconnects (action $A6$).

In the second scenario, say $S2$, the attacker intercepts the traffic
(action $A7$) exchanged between the IMD and the programmer, when
a physician was consulting the IMD configuration. Later, he replays
the access request (action $A8$), and leads the IMD to open a session
(action $A3$). Starting from the obtained state, the attacker executes
actions $A4$, $A5$, and $A6$ as in the scenario $S1$.

For the sake of space, we have not included the formal description
of each one of the actions part of the scenarios, nor provided the
description of states separating two actions in the scenario. However,
it can be easily understood that each executed action prepares for
the subsequent actions, and an action is executed from the state where
it is enabled. Moreover, as both of the scenarios contain malicious
actions, they could be considered as malicious scenarios.

Among the actions described above, actions $A1$, $A2$, $A7$, and
$A8$ are unobservable by the IMD as they target network resources,
and therefore not included in the collected evidences. The same is
true for action $A4$ which is legitimate and does not look to be
sensitive. Therefore, $obs(S1)=obs(S2)=\langle A3,A5,A6\rangle$.
That observation corresponds exactly to events $T1$, $T2$, and $T3$
available in the collected technical evidence.

\paragraph*{Correlation of the reconstructed scenarios}

Two cause/effect correlations allows to prove that each one of the
potential technical attack scenarios is the cause of the occurrence
of the medical scenario. Therefore the death of the patient could
be considered as a consequence of a lethal attack on the IMD. The
first correlation is between action $A5$ (therapy modification) in
the technical scenario, and inappropriate IMD response provided by
$(M1)$ in the medical scenario. The second correlation is between
$A5$ and absence of IMD response provided by $(M2)$ in the medical
scenario. That correlation states that the settings modified by $A5$
make the IMD unable to respond appropriately to the occurred arrhythmia.

\section{Conclusion}

\label{sec:Conclusion}

In this paper, we were interested in postmortem digital investigation
of attacks on cardiac IMDs. First, we identified a set of lethal attack
scenarios targeting IMDs, and we proposed a methodology for the postmortem
investigation analysis, which reconciles in the same framework conclusions
derived by technical investigators and pathologists. Second, an inference
system is described to provide an automated reconstruction of medical
scenarios staring from the medical evidence collected from the IMD
under investigation. These medical scenarios allow to identify and
understand heart and device anomalies that contributed to a patient
death. Third, we proposed a formal investigation technique for the
reconstruction of potential attack scenarios on IMDs, using an investigation
library and a set of collected technical evidence. A correlations
is performed to prove the causality between the reconstructed technical
and medical scenarios.

In future work, we will extend the IMD architecture to support the
generation and protection of a rich set of evidential data, and the
long-term energy-aware storage of these data. Another perspective
would consist in extending the proposed system to be tolerant to anti-forensic
attacks.

\nocite{*}
\bibliographystyle{eptcs}
\bibliography{references}

\end{document}